# Precision Measurement of the Anomalous Magnetic Moment of the Muon

William Morse[1] for the g-2 collaboration[*]
[1]*Brookhaven National Lab*

A precision measurement of the anomalous magnetic moment (g-2) of the muon has been made by the E821 collaboration at BNL. This paper is a written version of a "hot topics" talk at ICAP2002 (Cambridge, MA). The principles of the experiment are discussed, especially the beam dynamics aspects.

## 1 Introduction

The magnetic moment associated with orbital angular momentum is:

$$\mu_L = e\mathbf{L}/2mc$$

The magnetic moment associated with spin angular momentum is:

$$\mu_s = ge\mathbf{S}/2mc$$

The Dirac equation predicted $g = 2$ for a spin 1/2 particle without substructure. Measurements of the $g$ value of the electron were consistent with 2. However, the proton $g$ value was measured to be $g_p = 5.8$. This turned out to be a harbinger of new physics and was finally explained, along with the $g$ value of the neutron, $g_n = -3.8$, in the 1960s by the quark model. Attempts by Oppenheimer and others to calculate the first order correction to $g= 2$ for a structure-less particle gave infinity.

In 1947, the year the author was born, a more precise measurement of the electron g-value was made: $g_e = 2.002$. The anomalous magnetic moment was defined to be:

$$a = (g\text{-}2)/2$$

Several bright young theorists were able to calculate the first order correction to $g$=2 for a point-like spin 1/2 particle: $a = \alpha/2\pi$. Once again measurements of the anomalous magnetic moment turned out to be a harbinger of new physics: this time it was QED. Now the race was on between the theorists and the experimentalists to test our understanding with precision measurements of lepton anomalous magnetic moments compared to the calculations. The race continues even to this day.

In the 1970s, the weak interaction was unified with QED. The dominant electro-weak Feynman diagram is shown below.

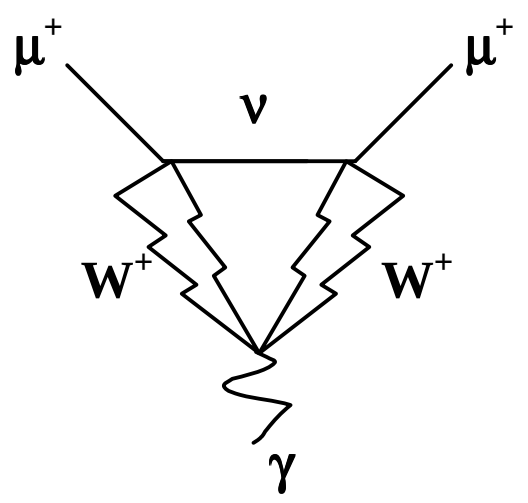

Figure 1: Feynman diagram with the largest contribution to $a_{EW}$.

This diagram has a contribution to the electron anomaly relative to $\alpha/2\pi$ of 0.08 parts per billion. This is much smaller than the experimental uncertainty in $\alpha$. However, the contribution to the muon anomaly relative to $\alpha/2\pi$ is 3 parts per million. This is because the sensitivity to a large mass scale $\Lambda$ much greater than the mass of the lepton $m_l$, is proportional to:

$$da_l = (m_l/\Lambda)^2$$

Thus the muon is perfect for searching for new mass scales [1]. In fact, it was pointed out that if nature has chosen supersymmetry to keep the weak mass scale from exploding, it may be that the supersymmetric version of the above Feynman diagram is even larger than the standard model version [2].

The price we pay for this lovely large muon mass is that the muon decays: $c\tau_\mu = 660m$. However, the decay violates parity maximally, giving us a beautiful way of continually monitoring the muon spin. The lepton lifetime is proportional to the inverse fifth power of the mass, making the tau lepton much too heavy for this type of experiment. In fact, the muon mass is about perfect for this type of experiment, perhaps partially answering Rabi's question when informed of the existence of the muon: "Who ordered that?"

# 2 The Principle of the Measurement

The experimental observable for parity violation is $\boldsymbol{S}\bullet\boldsymbol{P}$, so we want to keep track of the muon's spin relative to it's momentum. When a muon is stored in a storage ring with uniform magnetic field and $\boldsymbol{B}\bullet\boldsymbol{P}=0$, the cyclotron frequency is:

$$\omega_c = e\boldsymbol{B}/mc\gamma$$

This is the rate at which the momentum rotates. The rate at which the spin rotates is:

$$\omega_s = e\boldsymbol{B}/mc\gamma + ea\boldsymbol{B}/mc$$

where the above equation has included the Thomas precession, which takes into account that the muon is in a rotating reference frame and not an inertial frame. The difference between the spin and momentum rotation rates is just:

$$\boldsymbol{\omega_a} = \boldsymbol{\omega_s} - \boldsymbol{\omega_c} = ea\boldsymbol{B}/mc$$

Another miracle! The above equation contains no γ, but only terms which can be measured to high accuracy. In our experiment, we use a very uniform magnetic field with electrostatic quadrupole focusing. The anomaly equation for magnetic and electric fields [3] is:

$$\boldsymbol{\omega_a} = (e/mc)\,[a\boldsymbol{B} - \{a - 1/(\gamma^2 - 1)\}(\boldsymbol{\beta} \times \boldsymbol{E})]$$

We use the so-called magic momentum:

$$p_m = m_\mu c/\sqrt{a} = 3.09 GeV/c$$

where the coefficient in front of the electric field term is close to zero. The magic momentum can be understood qualitatively as follows. As $\beta \rightarrow 1$, either an electric or magnetic field in the lab looks like a magnetic field in the muon rest frame, and thus the effect on the magnetic moment is greater than on momentum by the anomaly. However as $\gamma \rightarrow 1$, the electric field has a much larger effect on the momentum compared to the magnetic moment. The magic momentum is where the electric field has the same effect on both the magnetic moment and the momentum.

# 3 Some Beam Dynamics

The field index [4] is proportional to the electric quadrupole gradient and is defined by:

$$n = B/(R\beta)\ \ \partial E_V/\partial y$$

The electric field focuses vertically, but de-focuses horizontally, ie. for positive muons, the top and bottom electrodes are positive, while the side electrodes are negative. The stored muons undergo simple harmonic motion in the vertical plane:

$$y = Y \cos(Q_V\, s/R + \varphi_y)$$

For continuous quadrupole coverage, the vertical tune $Q_v = \sqrt{n}$. This is a good approximation for our ring, although exact calculations are used to set the high voltage. For the horizontal (actually radial $x = \rho - R_0$, where $R_0$ = 7.11m) plane:

$$x = x_e + X \cos(Q_H\, s/R + \varphi_x)$$

where

$$x_e = (p-p_m)R/[p_m(1-n)]$$

and the horizontal tune $Q_h = \sqrt{(1-n)}$, reflecting the focusing effect of the magnetic field and the de-focusing effect of the electric field. As usual, one wants to avoid beam dynamics resonances where an integer times the horizontal tune plus an integer times the vertical tune equals an integer. Resonance lines up to fifth order with the working line $Q_h^2 + Q_v^2 = 1$ are shown in Fig. 2, along with the tune values used during the 2000 and 2001 runs. Short runs taken on resonances showed increased muon losses. We expect slightly higher muon losses for the 2001 running conditions.

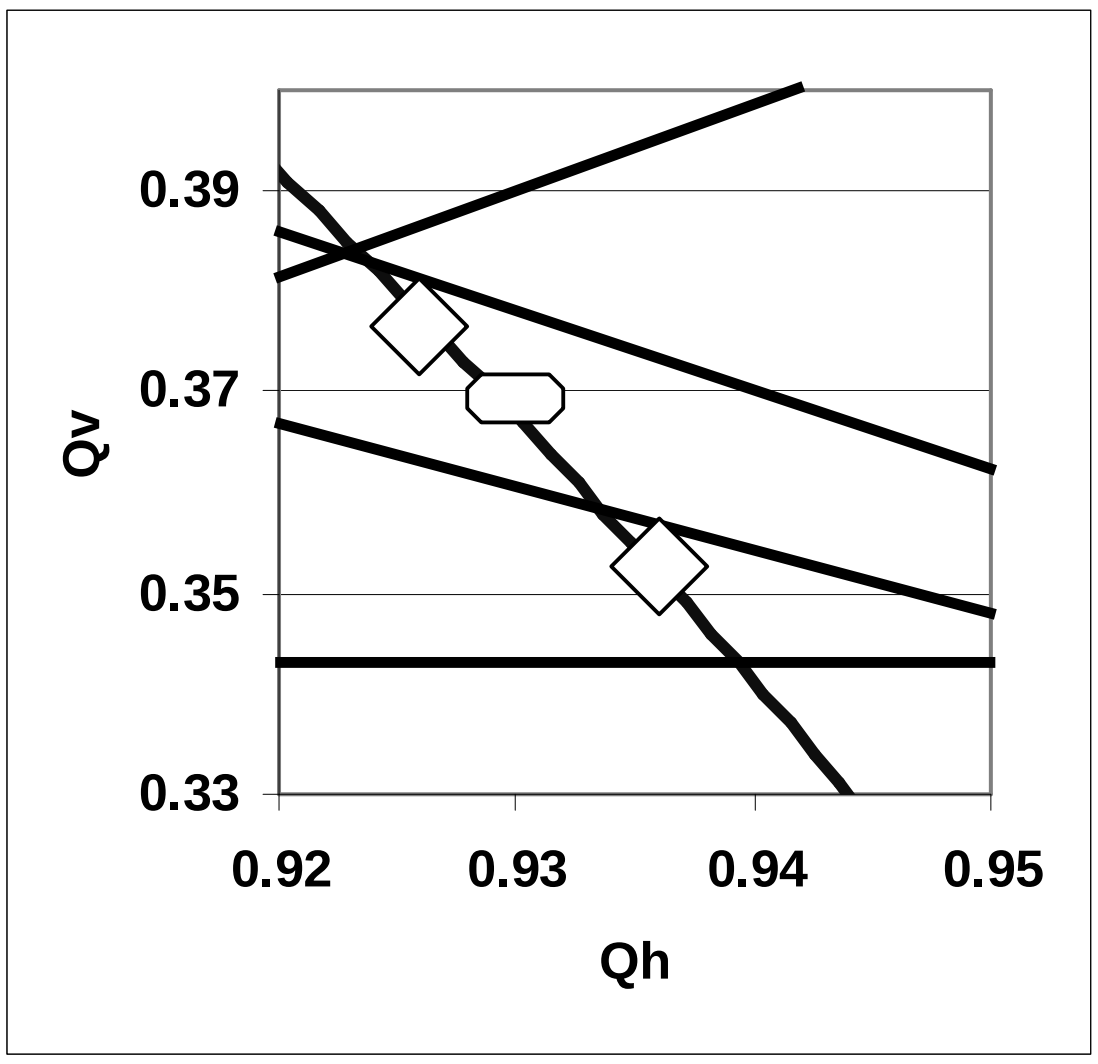

Figure 2: Resonance lines up to fifth order. The 2000 run quad settings are represented by an oval. The 2001 run quad settings are represented by diamonds.

An on-line measure of the number of stored muons vs. electrostatic quadrupole high voltage is shown in Fig. 3. The 2000 run high voltage settings were ±24KV, which corresponds to $n$ = 0.137. As the high voltage is increased, the vertical phase space increases, while the horizontal and momentum phase space decrease. The dependence of the number of stored muons on the quadrupole high voltage is in agreement with calculations. High voltage values up to ±25KV gave reliable quadrupole operation.

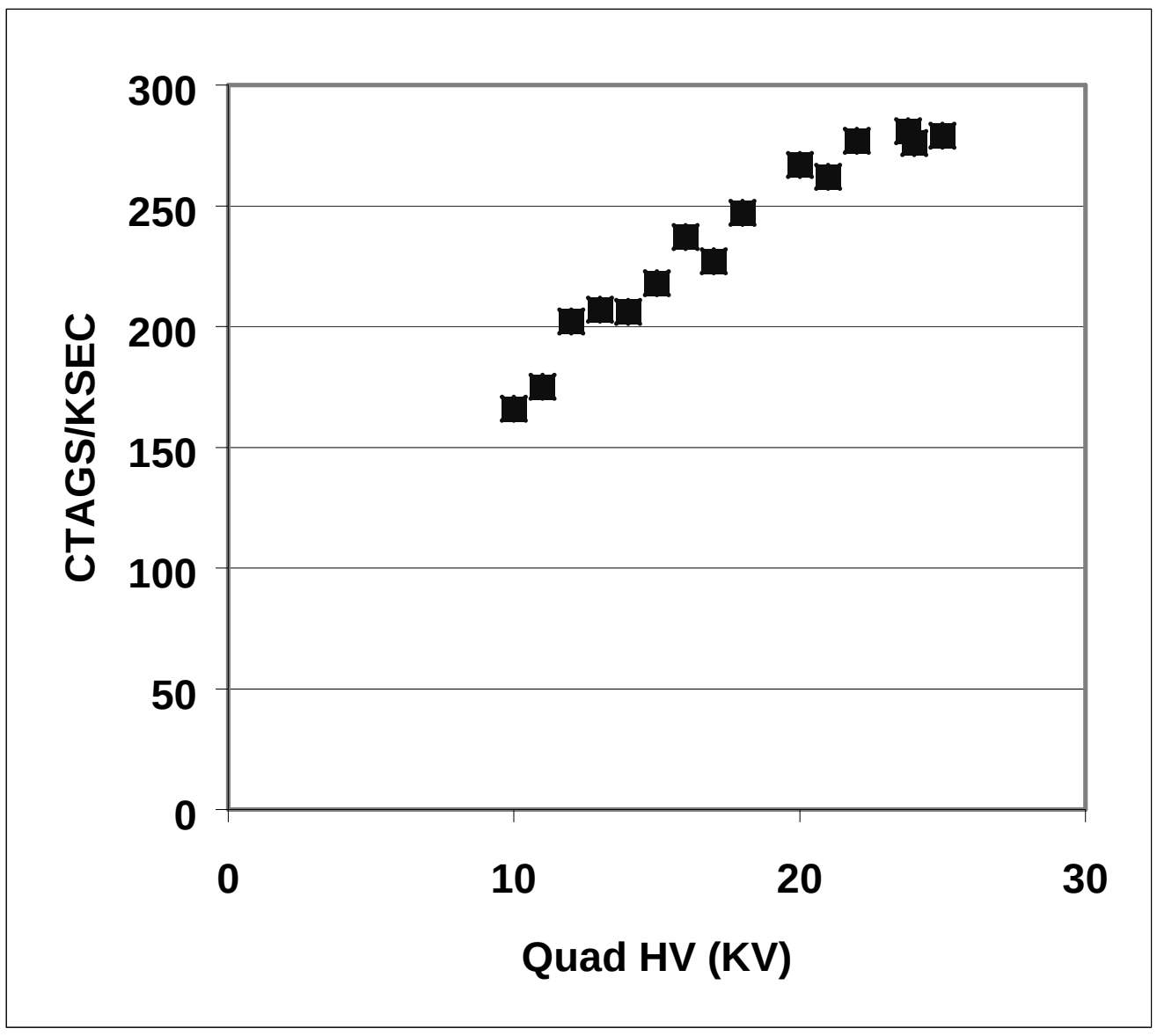

Figure 3: Normalized number of stored muons for short runs varying the quadrupole high voltage.

However, before we store muons we have to inject them. An inflector magnet [5] provides a field free channel of approximately 18mm(H) × 56mm(V) which brings the beam parallel to the central equilibrium orbit of $R_0$ = 7.11mm, but 7.6cm outside. The inflector channel cross-section is considerably less than the storage region cross-section of 90mm diameter. Previous measurements [6] used pion injection where the $\pi \rightarrow \mu\nu$ decay provides the kick to store the muons. Our experiment uses a fast kicker magnet [7] which provides an approximately 10mrad kick ninety degrees from the inflector magnet. This results in approximately an order of magnitude more useful stored muons when compared to pion injection. However, unlike with pion injection, the storage ring phase space is not filled, especially for horizontal betatron oscillations, due primarily to the 18mm vs. 90mm mismatch. This results initially in coherent betatron motion at the cyclotron frequency times the horizontal tune. The observed frequency at one location is $(1\text{-}Q_h)f_c$ due to aliasing. This is shown in Fig. 4 for two detectors opposite each other in the ring. The phase change between their observations is $\pi$. Specifically, Fig. 4 shows betatron motion $M = \cos(Q_H\, 2\pi t/\tau_c + \varphi_x)$ versus time, where $\tau_c = 150\eta s$, $Q_H = 3/4$ and aliasing is observed at $1\text{-}\ Q_H = 1/4$, to illustrate the effect. During run conditions, $Q_H \approx 0.93$ and $1\text{-}\ Q_H \approx 0.07$.

For our ring, $f_c$ = 6.70MHz and $f_a$ = 229KHz. The n values were chosen to be far from spin resonances: $K \pm LQ_H \pm MQ_V = f_a/f_c = \gamma a$. Short runs were not taken on the spin resonances, since their effects would not be observable even for long runs, due to the very uniform magnetic field.

## 4 Effect of $(1-Q_h)f_c$ on $\omega_a$ fits

Due to acceptance effects, the detectors clearly see this frequency in the decay spectra. For each detector, there is a modulation of the number of decays with this frequency of about 1%, a modulation of the parity violation asymmetry of about 0.1% and the g-2 phase of about 1mrad, in agreement with simulations of the detector acceptance. The coherent betatron motion lifetime due to the dp/p of the beam and non-linear terms from the electrostatic quadrupoles is measured to be about 120μs, in agreement with beam dynamics calculations. The effect on the measurement of $\omega_a$ due to the neglect of the asymmetry and phase modulation was studied analytically, with simulation, and with systematic studies. Unfortunately, it is enhanced when the observational coherent motion frequency is close to twice $\omega_a$, which was the situation for the 1999 and 2000 data taking runs. Fig. 5 shows the effect on $\omega_a$ from simulation.

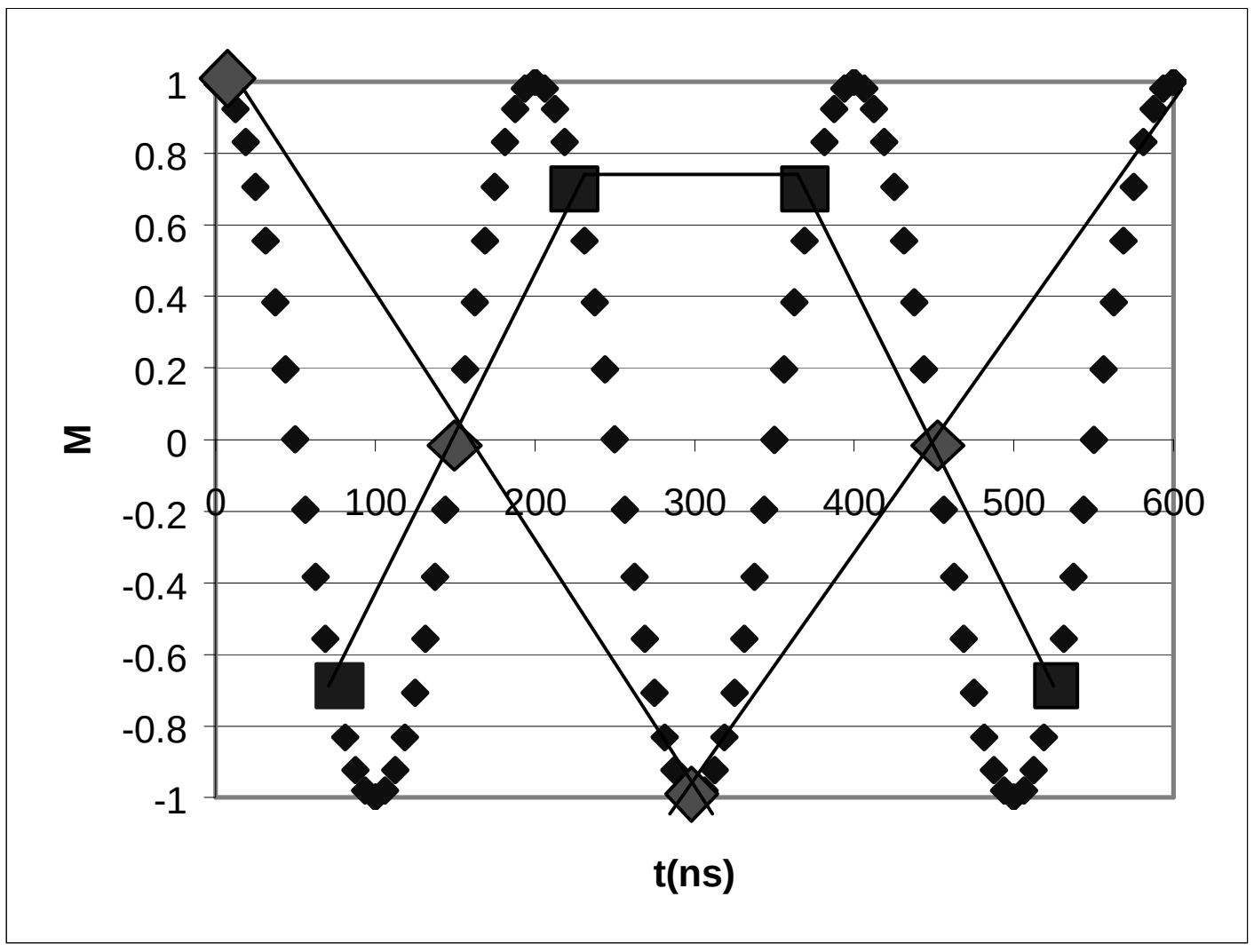

Figure 4: Observation of coherent betatron motion showing aliasing (see text).

Table I: Systematic uncertainties for the $\omega_a$ analysis.

| Source of errors | Size (ppm) |
|---|---|
| Coherent betatron motion | 0.21 |
| Pileup | 0.13 |
| Gain changes | 0.13 |
| Lost muons | 0.10 |
| Binning and fitting procedures | 0.06 |
| Others [8] | 0.06 |
| Total systematic error on $\omega_a$ | 0.31 |

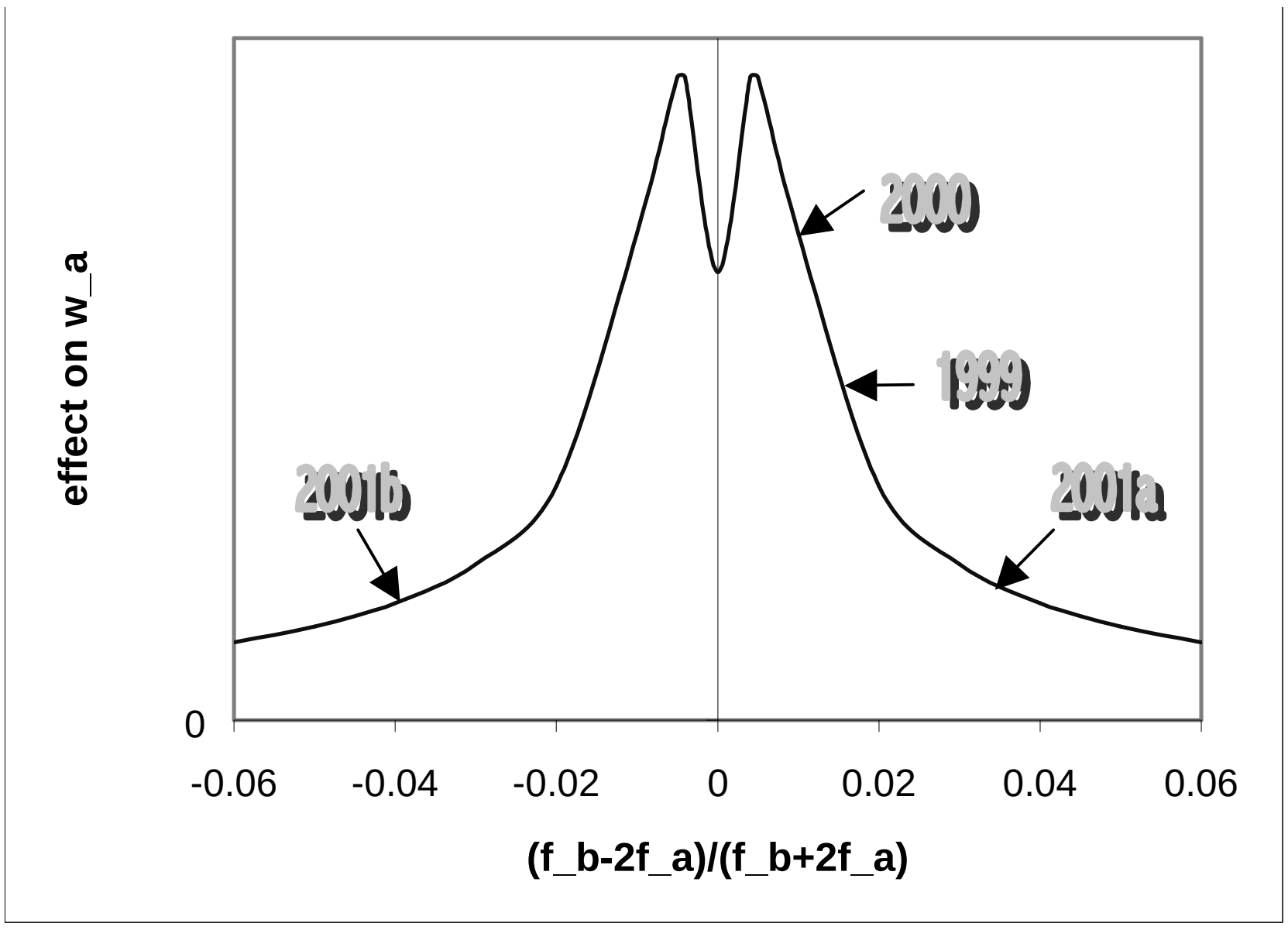

Figure 5: Simulated effect on $\omega_a$ of coherent betatron motion if the asymmetry and phase modulation are not included in the fit, as a function of the normalized difference between $f_b$ and $2f_a$.

# 5 Discussion

Numerous systematic studies were performed: neglecting the $\omega_b$ modulation terms in the fits to $\omega_a$ for individual detectors and summed data, including the terms individually and collectively in the fits, and fitting to $\omega_a$ for data strobed at the observed coherent betatron motion frequency. The estimated systematic error on $\omega_a$ for coherent betatron motion for the 2000 analysis is 0.21ppm, which is the largest single $\omega_a$ systematic error (see Table I). The total error for the analysis of the 2000 data sample is 0.7ppm, which includes the statistical error of 0.62ppm and the $\omega_p$ systematic error of 0.24ppm. The results and comparison with theory are given in ref. 9.

Now we will return to the analysis of the data taken in 2001. This consists of about three billion $\mu^-$ decays, to be compared to four billion $\mu^+$ decays from the 2000 run and one billion from the 1999 run. The coherent betatron motion systematic error is expected to be about a factor of three smaller, since the data were taken far from the observational resonance $\omega_b = 2\omega_a$. The PAC approved another $\mu^-$ run, but alas, the President's DOE budget zeroed out AGS running for FY03. Ten years ago Prof. Bromley from Yale gave a colloquium at BNL. The theme was the need to double the funding for the physical sciences over ten years. It's time.

*G.W. Bennett[2], B. Bousquet[9], H.N. Brown[2], G. Bunce[2], R.M. Carey[1], P. Cushman[9], G.T. Danby[2], P.T. Debevec[7], M. Deile[11], H. Deng[11], W. Deninger[7], S.K. Dhawan[11], V.P. Druzhinin[3], L. Duong[9], E. Efstathiadis[1], F.J.M. Farley[11], G.V. Fedotovich[3], S. Giron[9], F.E. Gray[7], D. Grigoriev[3], M. Grosse-Perdekamp[11], A. Grossmann[6], M.F. Hare[1], D.W. Hertzog[7], X. Huang[1], V.W. Hughes[11], M. Iwasaki[10], K. Jungmann[5], D. Kawall[11], B.I. Khazin[3], J. Kindem[9], F. Krienen[1], I. Kronkvist[9], A. Lam[1], R. Larsen[2], Y.Y. Lee[2], I. Logashenko[1,3], R. McNabb[9], W. Meng[2], J. Mi[2], J.P. Miller[1], W. Morse, D. Nikas[2], C.J.G. Onderwater[7], Y. Orlov[4], C.S. Ozben[2], J.M. Paley[1], Q. Peng[1], C.C. Polly[7], JPretz[11], RPrigl[2], G. zu Putlitz[6], T. Qian[9], S.I. Redin[3,11], O. Rind[1], B.L. Roberts[1], N. Ryskulov[3], P. Shagin[9], Y.K. Semertzidis[2], Yu.M. Shatunov[3], E.P. Sichtermann[11], E. Solodov[3], M. Sossong[7], A. Steinmetz[11], L.R. Sulak[1], A. Trofimov[1], D. Urner[7], P. von Walter[6], D. Warburton[2], and A. Yamamoto[8].

*[1]Boston University, [2]Brookhaven National Laboratory, [3]Budker Institute of Nuclear Physics, [4]Cornell University, [5]Kernfysisch Versneller Instituut, [6]Physikalisches Institut der Universit at Heidelberg, [7]University of Illinois, [8]KEK, [9]University of Minnesota, [10]Tokyo Institute of Technology, [11]Yale University*